# ADMIRE FRAMEWORK: DISTRIBUTED DATA MINING ON DATA GRID PLATFORMS

Nhien-An Le-Khac

*School of Computer Science & Informatics, University College Dublin*
*Belfield, Dublin 4, IRELAND*
*Email: an.lekhac@ucd.ie*

Tahar Kechadi

*School of Computer Science & Informatics, University College Dublin*
*Belfield, Dublin 4, IRELAND*
*Email: tahar.kechadi@ucd.ie*

Joe Carthy

*School of Computer Science & Informatics, University College Dublin*
*Belfield, Dublin 4, IRELAND*
*Email: joe.carthy@ucd.ie*

Abstract: In this paper, we present the ADMIRE architecture; a new framework for developing novel and innovative data mining techniques to deal with very large and distributed heterogeneous datasets in both commercial and academic applications. The main ADMIRE components are detailed as well as its interfaces allowing the user to efficiently develop and implement their data mining applications techniques on a Grid platform such as Globus ToolKit, DGET, etc.

## 1 INTRODUCTION

Distributed data mining (DDM) techniques have become necessary for large and multi-scenario datasets requiring resources, which are heterogeneous and distributed. There are two scenarios of datasets distribution: homogeneous and heterogeneous. Differences between them are presented in (Kargupta,2002). Actually, most of the existing research work on DDM is dedicated to the first scenario; there exists very little research in the literature about the second scenario.

Furthermore, today we have a deluge of data from not only science fields but also industry and commerce fields. Massive amounts of data that are being collected are often heterogeneous, geographically distributed and owned by different organisations. There are many challenges concerning not only DDM techniques but also the infrastructure that allows efficient and fast processing, reliability, quality of service, integration, and extraction of knowledge from this mass of data.

To face with large, graphically distributed, high dimensional, multi-owner, and heterogeneous datasets, Grid (Foster,2001) can be used as a data storage and computing platform to provide an effective computational support for distributed data mining applications. Because Grid is an integrated infrastructure that supports the sharing and coordinated use of resources in dynamic heterogeneous distributed environments. Actually, few projects have just been started using Grid as a platform for distributed data mining (Cannataro'a,2002)(Brezany,2003). However, they have only used some basic Grid services and most of them are based on Globus Tool Kit (Globus Tool Kit,website).

In this paper, we present the architecture of ADMIRE; a new framework based on Grid infrastructure for implementing DDM techniques. This framework is based on a Grid system dedicated to deal with data issues. Portability is key of this framework. It allows users to transparently exploit not only standard Grid architecture (Open Grid Services Architecture/Web Services Resource Framework-OGSA/WSRF) but also P2P/Grid architectures (DGET (Kechadi,2005)) for their data mining applications. Moreover, this new framework allows more dynamic and autonomous in the mining, integrating and processing phases. The rest of this paper organised as follows: section 2 deals with background and related projects; then we will discuss the ARMIRE architecture in section 3. Section 4 presents our preliminary evaluations of this architecture. Finally, we conclude on section 5.

## 2 BACKGROUND

In this section, we present firstly Grid architectures that ADMIRE is based on: OGSA/WSRF (OGSA,website) and DGET (Data Grid environment and Tools)(Kechadi,2005). We present also some common architectures for DDM systems: client/server based and agent-based as well as Grid data mining.

### 2.1 OGSA/WSRF

Actually, OGSA is a new-generation of Grid technologies. It becomes a standard for building service-oriented for applications that are based on Web services.

**Web services**

Traditional protocols for distributed computing such as Socket, Remote Procedure Call, Java RMI, CORBA (CORBA,website), etc. are not based on standards, so it is difficult to take up in heterogeneous environments. Today there are Web services which are open standards. For instance, XML-based middleware infrastructure can build and integrate applications in heterogeneous environments, independently of the platforms, programming languages or locations. It is based on Service Oriented Architecture Protocol (SOAP) that is a communication protocol for clients and servers to exchange messages in a XML format over a transport-level protocol HTTP. So Web services provide a promising platform for Grid systems.

**OGSA/WSRF**

OGSA is an information specification to define a common, standard and open architecture for Grid-based applications as well as for building service-oriented Grid systems. It is developed by Global Grid Forum and its first released was in June 2004. OGSA is based on Web-services and standardises almost Grid-services such as job and resource management, communication and security. These services are divided in two categories: core services and platform services. The former includes service creation, destruction, life cycle management, service registration, discovery and notification. The later deals with user authentication and authorisation, fault tolerance, job submission, monitoring and data access.

However, it does not specify how these interfaces should be implemented. This is provided by Open Grid Services Infrastructure (OGSI) (OGSI,website) that is the technical specification to specifically define how to implement the core Grid services as defined in OGSA in the context of Web services. It specifies exactly what needs to be implemented to conform to OGSA. One of the OGSI disadvantages is that it is too heavy with everything in one specification, and it does not work well with the existing Web service and XML tools. Actually, OGSI is being replaced by Web Services Resource Framework (WSRF) (WSRF,website) which is a set of Web services specifications released in January 2004. WSRF is a standard for modelling stateful resources with Web services. One of the advantages of WSRF is that it partitions the OGSI functionality into a set of specifications and it works well with Web services tools by comparison with OGSI. WSRF is now supported by Globus ToolKit 4 (Globus Tool Kit,website).

### 2.2 DGET

DGET (Kechadi,2005) is a Grid platform dedicated to large datasets manipulation and storing. Its structure is based on peer-to-peer communication system and entity-based architecture. The principal features of DGET are: transport independent communication, decentralised resource discovery, uniform resource interfaces, fine grained security, minimum administration, self-organisation and entity based system. A DGET entity is the abstraction of all logical or physical objects existing in the system. DGET communication model has been designed to operate without any centralized server as we can found in peer-to-peer systems. This feature is in contrast to many centralised grid system. Moreover, DGET also provides job migration that is not supported by some Grid platforms (Unicore, for example). Besides, DGET uses a sophisticated security mechanism based on Java security model.

### 2.3 DDM system architectures

Usually, there are two common architectures of DDM systems: client-server based and agent-based (Kargupta,2000). In the first one, they have proposed three-tier client-server architecture that includes client, data server and computing server. The client allows to create data mining tasks, visualisation of data, models. Data server deals with data access control, task coordination and data management. The computing server executes data mining services. This architecture is simple but it is not dynamic, autonomous and does not address to the scalability issue.

The second architecture aims to provide a scalable infrastructure for dynamic and autonomous mining over distributed data of large sizes. In this model, each data site has one or more agents for analysing local data and communication with agents at other sites during the mining. The locally mined knowledge will be exchanged among data sites to integrate a globally coherent knowledge.

Today, new DDM projects aim to mine data in a geographically distributed environment. They are based on Grid standards (e.g. OGSA) and platforms (e.g. Globus Toolkit) in order to hide the complexity of heterogeneous data and lower level details. So, their architectures are more and more sophisticated to articulate with Grid platforms as well as to supply a user-friendly interface for executing data mining tasks transparently. Our research also aims to this architecture.

## 2.4 Related Projects

This section presents two most recent DDM projects for heterogeneous data and platforms: Knowledge Grid (Cannataro'a,2002) (Cannataro'b,2002) and Grid Miner (Brezany,2003) (Brezany,2004). All of them use Globus Toolkit (Globus Tool Kit,website) as a Grid middleware.

**Knowledge Grid**

Knowledge Grid (KG) is a framework for implementing distributed knowledge discovery. This framework aims to deal with multi-owned, heterogeneous data. This project is developed by Cannataro el al. at University "Magna Graecia" of Catanzaro, Italy. The architecture of KG is composed of two layers: Core K-Grid and High level K-Grid.

The first layer includes Knowledge Directory service (KDS), Resource allocation and execution management service (RAEMS). KDS manages metadata: data sources, data mining (DM) software, results of computation, etc. that are saved in KDS repositories. There are three kinds of repositories: Knowledge Metadata Repository for storing data, software tool, coded information in XML; Knowledge Base Repository that stores all information about the knowledge discovered after parallel and distributed knowledge discovery (PDKD) computation. Knowledge Execution Plan Repository stores execution plans describing PDKD applications over the grid. RAEMS attempts to map an execution plan to available resource on the grid. This mapping must satisfy users, data and algorithms requirements as well as their constraints.

High level layer supplies four service groups used to build and execute PDKD computations: Data Access Service (DAS), Tools and Algorithms Access Service (TASS), Execution Plan Management Service (EPMS) and Results Presentation Service (RPS). The first service group is used for the search, selection, extraction, transformation, and delivery of data. The second one deals with the search, selection, download DM tools and algorithms. Generating a set of different possible execution plans is the responsible of the third group. The last one allows to generate, present and visualize the PDKD results.

The advantage of KG framework that it supports distributed data analysis and knowledge discovery and knowledge management services by integrating and completing the data grid services. However this framework was the non-OGSA-based approach.

**GridMiner**

GridMiner is an infrastructure for distributed data mining and data integration in Grid environments. This infrastructure is developed at Institue for Software Science, University of Vienna. GridMiner is a OGSA-based data mining approach. In this approach, distributed heterogeneous data must be integrated and mediated by using OGSA-DAI (DAI,website) before passing data mining phase. Therefore, they have divided data distribution in four data sources scenarios: single, federate horizontal partitioning, federate vertical partitioning and federate heterogeneity.

The structure of GridMiner consists of some elements: Service Factory (GMSF) for creating and managing services; Service Registry (GMSR) that is based on standard OGSA registry service; DataMining Service (GMDMS) that provides a set of data mining, data analysis algorithms; PreProcessing Service (GMPPS) for data cleaning, integration, handling missing data, etc.; Presentation Service (GMPRS) and Orchestration Service (GMOrchS) for handling complex and long-running jobs.

The advantages of GridMiner is that it is an integration of Data Mining and Grid computing. Moreover, it can take advances from OGSA. However, it depends on Globus ToolKit as well as OGSA-DAI for controlling data mining and other activities across Grid platforms. Besides, distributed heterogeneous data must be integrated before the mining process. This approach is not appropriate for complex heterogeneous scenarios.

## 3 ADMIRE ARCHITECTURE

ADMIRE architecture includes three main layers: interface, core and virtual data grid.

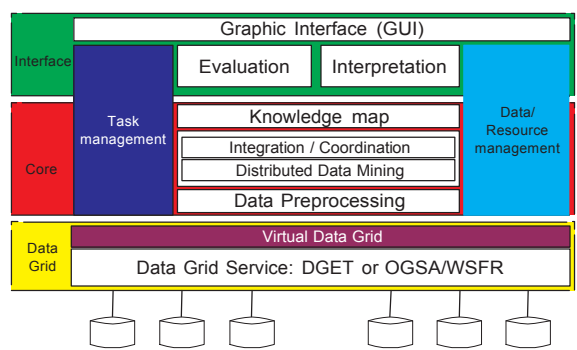

Figure 1: ADMIRE Software architecture.

## 3.1 Interface

The upper of this layer is a graphical user interface (GUI) allowing the development and the execution of DDM applications. By using this interface, users can build an DDM job including one or many tasks. A task contains either one of the DM techniques such as classification, association rules, clustering, prediction or other data operations such as data pre-processing, data distribution. Firstly, users choose a tasks and then they browse and choose resources that are represented by graphic objects, such as computing nodes, datasets, DM tools and algorithms correspondent to DM technique chosen. These resources are either on local site or distributed on different heterogeneity sites with heterogenous platforms. However, ADMIRE allows users to interact with them transparently at this level. The second step in the building of a DDM job is to establish links between tasks chosen, i.e. the execution order. By checking this order, ADMIRE system can detect independent tasks that can be executed concurrently. Furthermore, users can also use this interface to publish new DM tools and algorithms.

This layer allows to visualize, represent as well as to evaluate results of an DDM application too. The discovered knowledges will be represented in many defined forms such as graphical, geometric, etc. ADMIRE supports different visualization techniques which are applicable to data of certain types (discrete, continual, point, scalar or vector) and dimensions (1-D, 2-D, 3-D). It also supports the interactive visualization which allows users to view the DDM results in different perspectives such as layers, levels of detail and help them to understand these results better. Besides the GUI, there are four modules in this layer: DDM task management, Data/Resource management, interpretation and evaluation.

The first module spans both Interface and Core layers of ADMIRE. The part in the Interface layer of this module is responsible for mapping user requirements via selected DM tasks and their resources to an executing schema of tasks correspondent. Another role of this part is to check the coherence between DM tasks of this executing schema for a given DDM job. The purpose of this checking is, as mentioned above, to detect independent tasks and then this schema is refined to obtain an optimal execution. After verifying the executing schema, this module stores it in a task repository that will be used by the lower part of this task management module in the core layer to execute this DDM job.

The second module allows to browse necessary resources in a set of resources proposed by ADMIRE. This module manages the meta-data of all the available datasets and resources (computing nodes, DM algorithms and tools) published. The part in the Interface layer of this module is based on these meta-data that are stored in two repositories: datasets repository and resources repository to supply an appropriate set of resources depending on the given DM task. Data/Resource management module spans both Interface and Core layers of ADMIRE. The reason is that modules in the ADMIRE core layer also need to interact with data and resources to perform data mining tasks as well as integration tasks. In order to mask grid platform, data/resource management module is based on a data grid middleware, e.g. DGET(Kechadi,2005).

The third module is for interpreting DDM results to different ordered presentation forms. Integrating/mining result models from knowledge map module in the ADMIRE core layer is explained and evaluated.

The last module deals with evaluation the DDM results by providing different evaluation techniques. Of course, measuring the effectiveness or usefulness of these results is not always straightforward. This module also allows experienced users to add new tools or techniques to evaluate knowledge mined.

## 3.2 Core layer

The ADMIRE core layer is composed of three parts: knowledge discovery, task management and data/resource management.

The role of the first part is to mine the data, integrate the data and discover knowledge required. It is the centre part of this layer. This part contains three modules: data preprocessing; distributed data mining (DDM) with two sub components: local data mining (LDM) and integration/coordination; knowledge map.

The first module carries out locally data preprocessing of a given tasks such as data cleaning, data transformation, data reduction, data project, data standardisation, data density analysis, etc. These preprocessed data will be the input of the DDM module. Its LDM component performs locally data mining tasks. The specific characteristic of ADMIRE by compared with other current DDM system is that different mining algorithms will be used in a local DM task to deal with different kind of data. So, this component is responsible for executing and these algorithms. Then, local results will be integrated and/or coordinated by the second component of DDM module to produce a global model. Algorithms of data-preprocessing and data-mining are chosen from a set of pre-defined algorithms in the ADMIRE system. Moreover, users can publish new algorithms to increase the performance.

The results of local DM such as association rules, classification, and clustering etc. should be collected and analysed by domain knowledge. This is the role of the last module: knowledge map. This module will generate significant, interpretable rules, models and knowledge. Moreover, the knowledge map also controls all the data mining process by proposing different strategies mining as well as for integrating and coordinating to achieve the best performance.

The task management part plays an important role in ADMIRE framework. It manages all executing plan created from the interface layer. This part reads an executing schema from the task repository and then schedules and monitors the execution of tasks. According to the scheduling, this task management part carries out the resource allocation and then finds the best and appropriate mapping between resources and task requirements. This part is based on services supplied by Data Grid layer (e.g. DGET) in order to find the best mapping. Next, it will activate these tasks (local or distribution). Because ADMIRE framework is for DDM, this part is also responsible for the coordination of the distributed execution that is, it manages communication as well as synchronisation between tasks in the case of cooperation during the pre-processing, mining or integrating stage.

The role of data/resource management part is to facilitate the entire DDM process by giving an efficient control over remote resources in a distributed environment. This part creates, manages and updates information about resources in the dataset repository and the resource repository. The data/resource management part goes with the data grid layer to provide an transparent access to resources across heterogeneous platforms.

## 3.3 Data Grid layer

The upper part of this layer, called virtual data grid, is a portable layer for data grid environments. Actually, most of the Grid Data Mining Projects in literature are based on Globus ToolKit (GT) (Cannataro'a,2002)(Brezany,2003). The use of a set of Grid services provided by this middleware gives some benefits. For instance, the developer do not waste time for dealing with heterogeneous of organizations, platforms, data sources, etc.; software distributing is more easier because GT is the most widely used middleware in Grid community. However, as we mentioned above, this approach depends on GT with security overhead and GT's organisation of system topology and those services configuration are manual.

In order to make ADMIRE more portable, and more flexible with regards to many existing data grid platform developed today, we build this portable layer as an abstract of virtual Grid platform. It supplies a general services operations interface to upper layers. It homogenises different grid middleware by mapping of DM tasks from upper layer to grid services according to OGSA/WSRF standard or to entities (Kechadi,2005) in DGET model. For instance, ADMIRE is carried on DGET middleware. The portable layer implements two groups of entities: data and resource entities. The first group deals with data and meta-data used by upper layers and second one deals with resources used. The DGET system guarantees the transparent access of data and resource across a heterogeneous platform.

By using this portable layer, ADMIRE can be carried easily on many kind of Data Grid platforms such as GT and DGET.

## 4 EVALUATION

**Interface**

By using a graphic interface, ADMIRE aims to provide a user-friendly interaction with DDM users who can operate transparently with resources in distributed heterogeneous environment via graphic objects. The choosing of resources for each DM task in a DDM job makes more convenient for users and for systems in checking the consistency of an executing schema. Besides, the DM tools and algorithms are grouped into different groups depending on DM tasks (classification, association, clustering, pre-processing, data distribution, evaluation, etc.) chosen. This approach will help user to build a DDM application more easier than to do it in KG approach (Cannataro'a,2002) where users have to choose separately resources. Moreover, KG system has to check the consistency of a DM application built to avoid, e.g. an execution link between a node object and a dataset object.

Furthermore, users can also publish new DM tools and DM algorithms as well as evaluation algorithms and tools via this interface. It also allows users to control better the mining and evaluation process.

**Core layer**

In this layer, ADMIRE tries to use efficient distributed algorithms and techniques and facilitate seamless integration of distributed resources for complex problem solving. Furnishing dynamic, flexible and efficient DDM approaches and DDM algorithms in ADMIRE, users or system may choose different algorithms for the same DDM task according to the type or characteristics of data. This is an distinguish feature of ADMIRE by compare with current related project. This feature allows ADMIRE to deal with distributed and different level of heterogeneous of datasets.

By providing local preprocessing, ADMIRE could avoid the complexity in federation or global integration of heterogeneous data (or data schema) that normally need the intervention of experienced user as in GridMiner (Brezany,2003).

Data/resource management and task management modules are not only to support all of the data mining activities but also to support heterogeneous process. The coordination of these modules with data mining modules (preprocessing, DDM, Knowledge Map) makes ADMIRE system more robust than other knowledge grid frameworks that have not proposed any data mining approach.

**Data Grid layer**

As mentioned above, ADMIRE is based on this layer to cope with various Grid environments. By using OGSA/WSRF services, Data Grid platform layer allows ADMIRE to be ported on whatever Data Grid platform based on this standard which is chosen by almost recent Grid projects. For example, this feature allows users to port ADMIRE on an Globus Toolkit platform which is the most widely used grid middleware. Users do not have to waste time to re-install the grid platform for this new DDM framework.

Moreover, this layer makes ADMIRE to articulate well with DGET to benefit from the advantages of this data grid environment presented in (Kechadi,2005). Firstly, the dynamic, self-organizing topology of DGET allows ADMIRE to realize DM tasks on dynamic and autonomic platforms. Next, ADMIRE can deal with a very large and highly distributed datasets, e.g. medical data or meteorological data, by the decentralized resource discovery approach of DGET. This approach does not apply on any specialized servers. We can, therefore, avoid the bottleneck problem in intensive DDM tasks. Moreover, this feature allows DDM applications to execute effectively in integration and coordination between different sites. Thirdly, DGET approach does not require users to have individual user account on the resources and it uses XACML (Moses,2005) to define access control policies. This feature allows a very large number of users participating in DDM activities in ADMIRE. Last but not least, DM tasks in ADMIRE is more flexible in the distributed environment by supporting the strong migration provided in DGET. In addition, the implementation of ADMIRE on DGET platform is straightforward.

The virtual data grid layer is the only modification to carry ADMIRE on new Grid platform. It also makes our approach more portable.

## 5 CONCLUSION

In this paper we have proposed a new framework based on Grid environments to execute new distributed data mining techniques on very large and distributed heterogeneous datasets. The architecture and motivation for the design have been presented. We have also discussed some related projects and compared them with our approach.

The ADMIRE project is still in its early stage. So, we are currently developing a prototype for each layer of ADMIRE to evaluate the system features, test each layer as well as whole framework and building simulation and DDM test suites.